\documentclass[cameraready]{Interspeech}
\title{
FlashTTS: Fast Streaming TTS with MTP Acceleration and X-pred Mean Flow Distillation
}
\author[affiliation={1},orcid=0009-0008-1320-9387]{Hanke}{Xie}
\author[affiliation={1},orcid=0009-0001-3785-7126]{Xiaming}{Ren}
\author[affiliation={1},orcid=0009-0005-7381-0818]{Dake}{Guo}
\author[affiliation={1},orcid=0009-0007-9166-3596]{Ruonan}{You}
\author[affiliation={1},orcid=0009-0005-5732-0755]{Wenhao}{Li}
\author[affiliation={1},orcid=0009-0003-8727-2411]{Jingbin}{Hu}
\author[affiliation={1},orcid=0009-0001-6706-0572]{Guobin}{Ma}
\author[affiliation={1},orcid=0009-0004-6979-503X]{Huakang}{Chen}
\author[affiliation={2},]{Kejie}{Xu}
\author[affiliation={2},]{Rui}{Huang}
\author[affiliation={2},]{Weiguo}{Tan}
\author[affiliation={2},]{Xianrong}{Wang}
\author[affiliation={1},correspondingauthor,orcid=0000-0003-0771-9926]{Lei}{Xie}

\address{
$^{1}$ Audio, Speech and Language Processing Group (ASLP@NPU), School of Software, \\
Northwestern Polytechnical University, Xi’an, China \\
$^{2}$ Huawei Technologies Co., Ltd
}
\email{
{hkxie}@mail.nwpu.edu.cn, 
lxie@nwpu.edu.cn 
}


\usepackage{comment}
\usepackage[table]{xcolor}
\usepackage{booktabs}
\usepackage{multirow}
\usepackage{siunitx}
\definecolor{lightgreen}{RGB}{230,245,230}

\begin{document}

\maketitle


\begin{abstract}
% 现如今随着对话系统的快速发展，如今对LLM-based TTS提出了越来越高的要求。 为了降低现如今对话系统整体时延，以及实现real-time对话系统。现在如今对话系统的TTS有主要两大需求，低时延与输入输出流式。 为了解决这些问题（To address these 问题，我们提出了FlashTTS， ）  
Recent progress in speech dialogue systems requires Text-to-Speech (TTS) models to be faster and more responsive. Modern speech dialogue systems impose two primary requirements on TTS models: low latency and support for streaming inputs and outputs. However, most existing single-codebook LLM-based TTS methods rely on multi-stage pipelines that lack native streaming capabilities. These systems typically suffer from high end-to-end latency due to slow autoregressive prediction and multi-step flow matching. To address these limitations, we propose FlashTTS, an open-source and low-latency streaming TTS framework. FlashTTS introduces a lagged multi-track architecture that natively processes streaming text and speech inputs, thereby eliminating the need for sentence-level buffering. To accelerate acoustic generation, we integrate parallel Multi-Token Prediction (MTP) with an X-pred mean flow matching decoder. This configuration achieves high-fidelity token-to-mel generation in exactly two function evaluations (2-NFE). By jointly optimizing input processing and decoding efficiency, FlashTTS offers a practical foundation for real-time speech dialogue systems. Experiments show that FlashTTS substantially reduces First-Packet Latency to 325ms compared to robust streaming baselines, all while preserving strong zero-shot voice cloning and cross-lingual intelligibility. Speech samples are available.\footnote{\url{https://aslp-lab.github.io/flashtts_demo}} The model code and checkpoints will be released as open source\footnote{\url{https://github.com/ASLP-lab/FlashTTS}}.

% \footnote{\url{https://anonymous.4open.science/r/flashtts-00DB}}. 

\end{abstract}

\section{Introduction}
Building on the generative capacity of large language models (LLMs), modern text-to-speech (TTS) systems have reached a stage where synthesized speech is nearly indistinguishable from human speech. By imitating the timbre, prosody, and speaking style of reference audio, these systems achieve high naturalness and strong zero-shot voice cloning across arbitrary speakers, establishing LLM-based TTS as a core component of contemporary conversational and interactive systems.
% SPEAR~\cite{kharitonov2023speak}
To enable speech synthesis with LLMs, speech tokenization methods based on vector quantization (VQ)~\cite{van2017neural} or finite scalar quantization (FSQ)~\cite{mentzer2023finite} bridge continuous speech signals and discrete token sequences that can be effectively modeled by language models. VALL-E~\cite{wang2023neural} represents a pioneering system that leverages an LLM operating on discrete tokens with a residual vector quantization (RVQ) tokenizer~\cite{encodec}. In this design, tokens from the first codebook are predicted autoregressively, while tokens from the remaining codebooks are generated using non-autoregressive models. Building on this foundation, subsequent work has evolved into several paradigms, delayed autoregressive modeling~\cite{peng2024voicecraft, guo2024fireredtts}, and hierarchical architectures~\cite{liao2024fish, xie2025fireredtts, hu2026qwen3, liu2025quarkaudio} that combine large and small autoregressive models across codebooks.

\begin{figure}[t]
  \centering
  \includegraphics[width=1.0\linewidth]{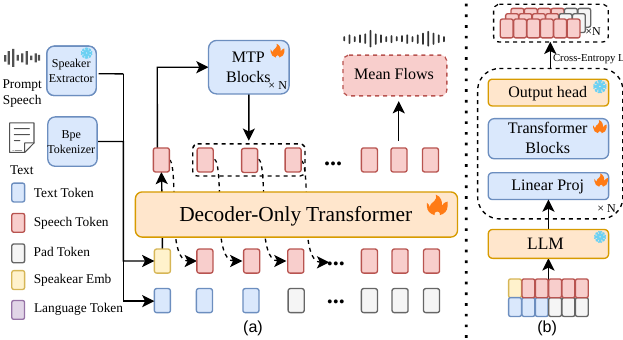}
  \caption{Architecture Overview of FlashTTS: (a) Stage 1 Training of the Stacked Inputs Track Structure. (b) Stage 2 MTP Training}
  \label{fig:flashtts}
  \vspace{-8mm}
\end{figure}

In contrast, an alternative paradigm predicts a single token stream that balances semantic and acoustic information, motivated by its closer alignment with intrinsic LLM modeling behavior. Early VQ-VAE-based approaches~\cite{betker2023_tortise_vqvae} rely on purely acoustic representations, which offer high reconstruction fidelity but lack explicit semantic supervision, making token prediction challenging. Later studies incorporate semantic information from self-supervised learning (SSL) representations, improving coherence and prosody~\cite{chen2025sac, ye2025llasa, wang2025spark}. These methods typically use dual-stream encoders to model semantic and acoustic features and fuse them into a unified token stream. From a language modeling perspective, a highly effective and increasingly prevalent strategy emphasizes the semantic capacity of LLMs by utilizing semantic tokens constrained by Automatic Speech Recognition (ASR) losses~\cite{casanova2024xtts, anastassiou2024seed, cosyvoice, cosyvoice2, cui2025glmttstechnicalreport}. These representations from ASR supervision provide remarkably stable, preserved linguistic information, which is easier for LLMs to predict. They serve as conditioning flow matching models for reconstructing timbre and prosody.

% {\color{blue}
% To reduce response delays, existing approaches often decrease model size, adopt chunk-based processing, or rely on system-level optimization.

Current speech dialogue systems, such as X-Talk~\cite{liu2025xtalk} and some commercial systems, typically rely on cascaded pipelines that combine LLM-based with flow matching speech synthesis. Deploying such interactive systems requires both extremely low latency and the capability to stream inputs and outputs.  However, conventional TTS systems do not natively support streaming input. Although recent work~\cite{xie2025fireredtts, xie2025soulx} introduces interleaved generation to reduce sentence-level waiting time, the overall end-to-end latency remains high. For output streaming, CosyVoice2 adopts chunk-aware causal flow matching together with a vocoder to accelerate waveform generation. This flow matching process commonly requires more than 10 sampling steps, which leads to high latency and instability during interleaved inference. More importantly, it does not address the main bottleneck caused by slow autoregressive token prediction, making it difficult for these systems to satisfy the real-time requirements of conversational scenarios.

To address these limitations, we propose FlashTTS, an open-source, low-latency TTS framework designed for streaming inputs in real-time speech dialogue systems. Based on the Qwen2-0.5B architecture, FlashTTS combines lagged multi-track streaming with parallel Multi-Token Prediction (MTP). This approach processes inputs incrementally and avoids the typical autoregressive bottleneck. For the acoustic decoder, we adopt an X-pred mean flow~\cite{geng2025mean,li2025back} strategy with block-level chunked attention, which directly predicts the mel-spectrogram, enabling high-fidelity token-to-mel generation in 2 inference steps. By minimizing both input and generation delays, FlashTTS seamlessly meets the real-time deployment needs of cascaded systems.

% }

% This design necessitates chunk-level segmentation, disrupting temporal continuity across audio segments and degrading perceived naturalness. In addition, conventional TTS systems lack native support for streaming input, preventing true simultaneous listening and speaking. The mismatch between token-level text generation and buffered TTS input further increases latency and fragments interaction, impairing real-time user experience.
% % 这部分修改一下，承上启下，上文先多夸一下asr token好处，稳定，利于预测，使用方案多.然后引出x talk glm4voice这种集联对话系统，里面需求主要是低时延和输入/输出流式。 然后现在方案cosyvoice用chunk flow matching解决输出流式，但还没有解决输入流式， 低时延
% While prior work emphasizes high-fidelity speech generation, streaming synthesis in interactive scenarios imposes stricter constraints on both naturalness and latency to preserve temporal smoothness and perceptual continuity. Some systems~\cite{xie2025fireredtts, xie2025soulx} adopt interleaved generation to reduce sentence-level waiting time, yet end-to-end latency remains high. CosyVoice2 introduces chunk-aware causal flow matching with a pretrained vocoder to accelerate waveform generation, but it does not address the core bottleneck of slow autoregressive token prediction and typically requires more than ten inference steps during decoding. Consequently, despite improved efficiency over conventional TTS systems, these methods still struggle to meet real-time requirements.

\section{FlashTTS}
\subsection{Model Architecture}
\begin{figure}[t]
  \centering
  \includegraphics[width=0.9\linewidth]{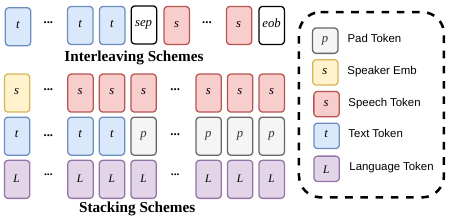}
  \caption{Comparison of different input schemes.}
  \label{fig:speech_production}
  \vspace{-4mm}
\end{figure}

In this section, we present FlashTTS, a low-latency TTS model explicitly designed for streaming input scenarios. As illustrated in Figure \ref{fig:flashtts}, the model architecture and training pipeline are structured into two primary stages. Stage 1 (Figure \ref{fig:flashtts}a) establishes the core generation pathway, wherein a trainable decoder-only transformer processes parallel input tracks composed of speech, text, and language representations. The resulting sequence is then reconstructed into an audio waveform via a mean flow matching module. Stage 2 (Figure \ref{fig:flashtts}b) incorporates Multiple Token Prediction (MTP) training. By leveraging a frozen LLM and supplementary trainable transformer blocks, this stage optimizes the system to predict multiple tokens simultaneously, effectively accelerating inference without compromising the generation quality.

\subsection{Streaming Sequence Schemes}

Unlike the conventional sequence organization of TTS systems, which typically formats data as interleaved or concatenated sequences of text and speech tokens, we adopt a stacked and lagged track structure to enable native streaming input. As depicted in Figure \ref{fig:speech_production}, the stacked configuration consists of three parallel information streams: a speech track, a text track, and a language track. The speech track is initialized with a speaker embedding and is subsequently followed by the generated speech tokens. Concurrently, the text track ingests input text tokens and applies padding tokens to maintain alignment once the text input concludes. Furthermore, the language track supplies continuous language conditioning throughout the generation process. By organizing the input into these parallel tracks, the system circumvents the necessity of buffering the complete text sequence prior to acoustic modeling. This design fundamentally aligns with the requirements of real-time interactions and substantially diminishes end-to-end latency.

\subsection{Multiple Token Prediction Acceleration}

Inspired by DeepSeek-V3~\cite{liu2024deepseekv3}, we incorporate the Multi-Token Prediction (MTP) framework, wherein each module comprises a linear projection layer followed by a Qwen2.5-Decoder block. Based on the empirical observation that the frame rate of text tokens typically ranges from 3 to 5 Hz, we restructure the conventional sequential architecture into a parallel formulation. As illustrated in Figure~\ref{fig:flashtts}b, parallel MTP modules directly ingest the final hidden states $\mathbf{h}_{0:t}^0$ generated by the backbone language model:
\begin{equation}
\mathbf{h}_{0:t}^{k} = \text{MTP}_k(\mathbf{h}_{0:t}^{0})
\label{eq:mtp_parallel}
\end{equation}
where $k \in \{1, 2, \dots, N - 1\}$ indicates the index of the module, and $0:t$ specifies the temporal span. Each module independently processes $\mathbf{h}_{0:t}^0$ to compute the respective output representations $\mathbf{h}_{0:t}^k$. These representations are then forwarded through a shared language model head to generate the distributions of future tokens. To mitigate the complexity of optimization, the parameters of the backbone remain frozen. The output probability distributions $\mathbf{S}^{k}$ are evaluated against the ground-truth speech tokens $\mathbf{G}$ via a cross-entropy loss function:
\begin{equation}
\mathcal{L}_{\text{MTP}} = \sum_{k=1}^{N-1} \mathcal{L}_{\text{CE}}(\mathbf{S}_{0:T-k-1}^{k}, \mathbf{G}_{k+1:T})
\end{equation}
where $T$ represents the total length of the sequence, and $\mathbf{G}_{k+1:T}$ is shifted to align with the objective of predicting the $(k+1)$-th future token. Furthermore, because the MTP modules are structurally lightweight, their direct token predictions can occasionally exhibit instability during inference. To ensure the quality of the synthesized speech, we draw inspiration from Llasa+~\cite{tian2025llasa+} and adopt a verification operation, that utilizes the robust probability distributions of the frozen backbone to validate the speculative tokens generated by the parallel modules.

\begin{table*}[htbp]
\small
\centering
\caption{Latency and quality comparison on the Minimax subset (zh, en, ja, ko). \textbf{FTL}: First-Token Latency (speech dialogue simulation), \textbf{TPP}: Tokenizer First-Packet Decode Latency, \textbf{FPL}: First-Packet Latency, \textbf{TPS}: Tokens Per Second, \textbf{RTF}: Real-Time Factor. ($\downarrow$ = lower is better, $\uparrow$ = higher is better).}
\label{tab:latency}
\setlength{\tabcolsep}{5pt}
\renewcommand{\arraystretch}{1.1}

\resizebox{0.85\textwidth}{!}{
\begin{tabular}{lccccccccc}
\toprule
\textbf{Model Configuration} & \textbf{TPS} $\uparrow$ & \textbf{FTL (ms)} $\downarrow$ & \textbf{LLM Lat. (ms)} $\downarrow$ & \textbf{TPP (ms)} $\downarrow$ & \textbf{FPL (ms)} $\downarrow$ & \textbf{RTF} $\downarrow$ & \textbf{WER} $\downarrow$ & \textbf{SIM} $\uparrow$ & \textbf{CMOS} $\uparrow$ \\
\midrule
\multicolumn{10}{l}{\textit{Baselines}} \\
CosyVoice2 (10-NFE) & 51 & 257 & 425 & 339 & 843 & 0.913 & 26.2 & 0.721 & 0.00 \\
\midrule
\multicolumn{10}{l}{\textit{FlashTTS (Proposed)}} \\
Stage 1 (2-NFE) & 50 & \textbf{60} & 246 & \textbf{123} & 377 & 0.793 & 18.0 & 0.702 & 0.08 \\
Stage 1 (3-NFE) & 50 & \textbf{60} & 246 & 198 & 413 & 0.824 & \underline{17.2} & 0.711 & 0.15 \\
MTP-3 (2-NFE) & \underline{73} & \underline{62} & \textbf{195} & \textbf{123} & \textbf{325} & \underline{0.632} & 18.8 & 0.695 & 0.05 \\
MTP-3 (3-NFE) & 72 & \underline{62} & 215 & \underline{165} & 366 & 0.702 & 17.5 & \underline{0.714} & \textbf{0.12} \\
MTP-5 (2-NFE) & \textbf{75} & \underline{62} & \underline{206} & \textbf{123} & \underline{328} & \textbf{0.621} & 20.8 & 0.668 & -0.08 \\
\bottomrule
\end{tabular}
}
\vspace{-5mm}
\end{table*}

\subsection{X-pred Mean Flow Distillation}
Drawing inspiration from recent advancements in generation methodologies, such as Mean Flow~\cite{geng2025mean} and JIT~\cite{li2025back}, we integrate the theoretical principles of Mean Flow~\cite{geng2025mean,ma2025meanvc,wang2025intmeanflow} with the explicit data prediction framework of JIT~\cite{li2025back} to facilitate high-quality synthesis with a minimal number of neural function evaluations (e.g., 2-NFE). In high-dimensional spaces, directly predicting velocity fields often leads to optimization difficulties. Consequently, the proposed formulation parameterizes the neural network to explicitly predict the clean mel-spectrogram $x$. The average velocity $u$ is subsequently derived from this data prediction. Furthermore, to explicitly accommodate the requirements of streaming audio generation, we incorporate a block-wise attention mechanism~\cite{guo2025streamflow}, seamlessly adapting the architecture to support native real-time streaming output.

Given a temporal interval $[r, t]$, the average velocity along the trajectory of the ordinary differential equation (ODE) is defined as:
\begin{equation}
u(z_t, r, t) \triangleq \frac{1}{t-r} \int_r^t v(z_\tau, \tau)\, d\tau.
\end{equation}
Differentiating with respect to $t$ yields the mean flow identity:
\begin{equation}
u(z_t, r, t) = v(z_t, t) - (t - r)\frac{d}{dt}u(z_t, r, t),
\end{equation}
where the total derivative is expanded as $\frac{d}{dt}u = v(z_t, t)\partial_z u + \partial_t u$. By substituting the marginal velocity with the conditional velocity $v_t = \epsilon - x$, the training target is formulated as $u_{\text{tgt}} = v_t - (t-r)(v_t \partial_z u_\theta + \partial_t u_\theta)$.

To optimize mel-spectrogram prediction, the neural network outputs the prediction $\hat{x}_\theta(z_t, r, t) = f_\theta(z_t, r, t)$. The estimated mean velocity $\hat{u}_\theta(z_t, r, t)$ is then analytically derived from $\hat{x}_\theta$:
\begin{equation}
\hat{u}_\theta(z_t, r, t) = \frac{1}{t} \big(z_t - \hat{x}_\theta(z_t, r, t)\big).
\end{equation}
The objective function is constructed to minimize the discrepancy between this derived average velocity $\hat{u}_\theta$ and the distillation target $u_{\text{tgt}}$:
\begin{equation}
\mathcal{L}_{\text{MF}}(\theta) = \mathbb{E}_{t,r,x,\epsilon}\left[ \left\lVert \hat{u}_\theta(z_t,r,t) - \operatorname{sg}(u_{\text{tgt}}) \right\rVert^2 \right],
\end{equation}
where $\operatorname{sg}(\cdot)$ denotes the stop-gradient operation. At the boundary condition $t = r$, this objective reduces to the standard conditional flow matching (CFM) loss. During the inference phase with 1-NFE sampling, the clean data sample is recovered as $z_0 = z_1 - \hat{u}_\theta(z_1, 0, 1)$, with the initial latent state sampled as $z_1 \sim p_{\text{prior}}(\epsilon)$.

\section{Experimental Setup}

\subsection{Datasets}
FlashTTS is trained on approximately 300{,}000 hours of open-source speech data, comprising Emilia, Emilia-Yodas~\cite{he2024emilia,ma2024wenetspeech4tts,kang2024libriheavy}, LibriHeavy, and WenetSpeech4TTS. We evaluate zero-shot performance on the Seed-TTS\footnote{https://github.com/BytedanceSpeech/seed-tts-eval} and MiniMax\footnote{https://huggingface.co/datasets/MiniMaxAI/TTS-Multilingual-Test-Set} multilingual test sets.

\subsection{Model Details}

\noindent\textbf{Model Configuration.} FlashTTS is built on the Qwen2.5-0.5B backbone~\cite{qwen}, with a hidden size of 896, 24 layers, 14 attention heads, and a feed-forward dimension of 4864. Each MTP module introduces an additional Qwen2.5-style decoder layer with the same architecture. The X-pred MeanFlow component adopts a 16-layer Diffusion Transformer with a hidden dimension of 768, approximately 159.25M parameters, followed by a 50M-parameter HiFi-GAN 24 kHz vocoder.

\noindent\textbf{Training Setup.} Training proceeds in two stages. In Stage 1, the full backbone is optimized using a dynamic frame-based batch size of 40{,}000 on 8 A100 GPUs with AdamW. The peak learning rate is $1 \times 10^{-4}$ with 20k warmup steps, followed by cosine decay over 1M training steps. In Stage 2, the backbone parameters are frozen, and only the MTP modules are trained on 4 A100 GPUs with a peak learning rate of $5 \times 10^{-5}$. At the same time, the X-pred MeanFlow model is distilled from a pretrained conditional flow matching teacher using 8 RTX 4090 GPUs, with a batch size of 2{,}000, gradient accumulation of 2, and a peak learning rate of $7 \times 10^{-5}$.

\noindent\textbf{Baselines.} Due to ASR tokens exhibiting high stability that facilitates language model prediction, we adopt S3Tokenizer v2 for speech tokenization. To ensure a fair token-level comparison under this scheme, we select CosyVoice2 (0.5B) as our primary baseline due to its comparable parameter scale. Furthermore, we include several other mainstream TTS models to provide a broader evaluation.

\subsection{Evaluation Metrics}

Speech quality and intelligibility are evaluated using WER, CER, SIM, and CMOS. Streaming efficiency is measured by Speedup Ratio, Real-Time Factor (RTF), Tokens Per Second (TPS), and First-Packet Latency (FPL). To assess subjective perceptual quality, we conduct Comparative Mean Opinion Score (CMOS) evaluations, each reported with 95\% confidence intervals. We randomly sample 100 test pairs and recruit 30 listeners for the listening tests. WER and CER are computed using Paraformer-zh for Chinese and Whisper-large-v3 for other languages\footnote{https://github.com/modelscope/FunASR}. Speaker similarity (SIM) is calculated as the cosine similarity between generated and reference speaker embeddings extracted from a fine-tuned WavLM-large model.

\section{Experimental Results}

\begin{table*}[t]
\centering
\caption{Objective evaluation metrics on the multilingual test set. 
Lower WER and higher SIM indicate better performance. 
``--'' denotes unsupported languages. \textbf{Bold} indicates the best result across all models, and \underline{underline} indicates the best result among open-source systems. * represents Commercial Model.}
\label{tab:multilingual_results}
\setlength{\tabcolsep}{3.4pt}
\renewcommand{\arraystretch}{1.15}

\begin{tabular}{l
S[table-format=2.2] S[table-format=1.3]
S[table-format=2.2] S[table-format=1.3]
S[table-format=2.2] S[table-format=1.3]
S[table-format=2.2] S[table-format=1.3]
S[table-format=2.2] S[table-format=1.3]
S[table-format=2.2] S[table-format=1.3]}
\toprule
\multirow{2}{*}{Model} 
& \multicolumn{2}{c}{Chinese} 
& \multicolumn{2}{c}{English} 
& \multicolumn{2}{c}{Japanese} 
& \multicolumn{2}{c}{Korean} 
& \multicolumn{2}{c}{French} 
& \multicolumn{2}{c}{German} \\
\cmidrule(lr){2-3}\cmidrule(lr){4-5}\cmidrule(lr){6-7}
\cmidrule(lr){8-9}\cmidrule(lr){10-11}\cmidrule(lr){12-13}
& {WER$\downarrow$} & {SIM$\uparrow$} 
& {WER$\downarrow$} & {SIM$\uparrow$} 
& {WER$\downarrow$} & {SIM$\uparrow$} 
& {WER$\downarrow$} & {SIM$\uparrow$} 
& {WER$\downarrow$} & {SIM$\uparrow$} 
& {WER$\downarrow$} & {SIM$\uparrow$} \\
\midrule

MiniMax{*} 
& {2.25} & {\textbf{0.780}}
& {\textbf{2.16}} & {0.756}
& {\textbf{3.52}} & {0.776}
& {\textbf{1.75}} & {0.776}
& {\textbf{4.10}} & {\textbf{0.628}}
& {1.91} & {\textbf{0.733}} \\

ElevenLabs{*} 
& {16.03} & {0.677}
& {2.34}  & {0.613}
& {10.65} & {0.738}
& {1.87}  & {0.700}
& {5.22}  & {0.535}
& {\textbf{0.57}} & {0.614} \\

\midrule

CosyVoice2 
& {1.22} & {\underline{0.773}}
& {3.44} & {\textbf{0.765}}
& {\underline{7.94}} & {\textbf{0.803}}
& {16.68} & {\textbf{0.778}}
& \multicolumn{2}{c}{--} 
& \multicolumn{2}{c}{--} \\

FlashTTS 
& {\textbf{1.08}} & {0.743}
& {\underline{3.02}} & {0.662}
& {10.59} & {0.751}
& {\underline{3.49}} & {0.734}
& {\underline{8.62}} & {\underline{0.564}}
& {\underline{9.96}} & {\underline{0.672}} \\

\bottomrule
\end{tabular}
\vspace{-4mm}
\end{table*}

\begin{table}[t]
    \centering
    \caption{TTS performance on Seed test sets (test-zh for Chinese, test-en for English). \textbf{Bold} indicates the best overall result.}
    \label{tab:seedtestset}
    \setlength{\tabcolsep}{4pt} 
    \small 

    \begin{tabular}{lcccc}
        \toprule
        \multirow{2}{*}{Model} & 
        \multicolumn{2}{c}{test-zh} & 
        \multicolumn{2}{c}{test-en} \\
        \cmidrule(lr){2-3} \cmidrule(lr){4-5}
        & CER$\downarrow$ & SIM$\uparrow$ & WER$\downarrow$ & SIM$\uparrow$ \\
        \midrule
        Seed-TTS~\cite{anastassiou2024seed} & \textbf{1.12} & \textbf{0.796} & 2.25 & \textbf{0.762} \\
        MaskGCT~\cite{wang2024maskgct}      & 2.27 & 0.774 & 2.62 & 0.714 \\
        F5-TTS~\cite{f5tts}                 & 1.56 & 0.741 & \textbf{1.83} & 0.647 \\
        Llasa-8B-250k~\cite{ye2025llasa}    & 1.59 & 0.684 & 2.97 & 0.574 \\
        Spark-TTS~\cite{wang2025spark}      & 1.20 & 0.672 & 1.98 & 0.584 \\
        CosyVoice2~\cite{cosyvoice2}        & 1.45 & 0.748 & 2.57 & 0.652 \\
        \midrule
        FlashTTS (stage 1) & 1.38 & 0.718 & 2.21 & 0.572 \\
        FlashTTS (stage 2) & 1.51 & 0.699 & 2.55 & 0.523 \\
        \bottomrule
    \end{tabular}
    % \vspace{-4mm}
\end{table}

\subsection{Latency and Quality Analysis}

All latency evaluations are conducted on the Minimax subset using a single NVIDIA RTX 4090 GPU, with the textual stream dynamically generated by an upstream Qwen2-7B model to simulate a real-world conversational pipeline. Crucially, all efficiency measurements are conducted without engineering optimizations to accurately reflect the raw architectural speed. While CosyVoice2 requires buffering five text tokens before synthesis, FlashTTS initiates generation with a single token, fundamentally reducing FTL. Table~\ref{tab:latency} details how this architecture balances efficiency and quality. Compared to the Stage 1 baseline, integrating three MTP branches significantly accelerates decoding through higher TPS and lower RTF, preserving competitive WER and SIM. Aggressively expanding to five MTP branches yields diminishing returns in speed and noticeably degrades acoustic fidelity, evidenced by a negative CMOS. Meanwhile, adjusting the MeanFlow NFE presents a clear physical trade-off: 3-NFE favors synthesis quality and achieves higher CMOS, whereas 2-NFE tightly constrains TPP and FPL while maintaining a positive subjective preference over the baseline. The MTP-3 (2-NFE) configuration therefore isolates the optimal operating point, delivering extreme real-time responsiveness for streaming speech dialogue scenarios without sacrificing core intelligibility or naturalness.

\subsection{Zero-Shot Speech Synthesis Performance}

We evaluate zero-shot voice cloning and multilingual generalization on a six-language MiniMax subset and public Seed test sets. Table~\ref{tab:multilingual_results} and Table~\ref{tab:seedtestset} demonstrate that FlashTTS provides robust intelligibility and extended language coverage. In cross-lingual scenarios, the model outperforms the CosyVoice2 baseline in English and Korean intelligibility while successfully supporting French and German. Although SIM moderately trails heavily parameterized offline models like Seed-TTS, FlashTTS maintains stable zero-shot cloning fidelity across languages. These results confirm that single-stream token modeling effectively aligns semantic and acoustic representations, accepting a reasonable trade-off in acoustic variance to drastically reduce end-to-end latency for real-time interactions.

\begin{table}[t]
\centering
\small 
\caption{Ablation study on FlashTTS components. The best and second-best results are highlighted in \textbf{bold} and \underline{underlined}, respectively.}
\label{table_ablation}
\resizebox{\linewidth}{!}{%
\begin{tabular}{lccc}
\toprule
Model                 & WER (\%) $\downarrow$ & SIM $\uparrow$   & Speed-Up Ratio (\%) $\uparrow$ \\ 
\midrule
CosyVoice2            & 2.21           & \textbf{0.743} & 0             \\
\midrule
FlashTTS              & \underline{2.17}  & 0.713 & \underline{49.23}  \\   
\quad w/o X-pred      & 2.28 & 0.691          & 12.53         \\ 
\quad w/o MTP         & \textbf{1.91}  & \underline{0.719} & 12.52         \\ 
\quad w/o Language ID & 3.42           & 0.702          & \textbf{49.28} \\
% \quad w/o Verification& 18.23          & 0.483          & \textbf{102.17}\\ 
\bottomrule
\end{tabular}%
}
\vspace{-4mm}
\end{table}

\subsection{Ablation Study}

Table~\ref{table_ablation} details the ablation of key FlashTTS components to isolate their impact on inference efficiency and acoustic quality. Both MTP and the X-pred strategy act as primary engines for acceleration; removing either module severely diminishes the speed-up ratio and limits real-time deployment viability. Furthermore, achieving this extreme acceleration requires explicit semantic guidance to maintain generation stability. For instance, dropping the language identification module noticeably degrades cross-lingual alignment and increases WER. Ultimately, the complete architecture strikes a necessary equilibrium where MTP and X-pred jointly maximize throughput, while explicit language conditioning tightly bounds acoustic variance during high-speed parallel decoding.

\section{Conclusion}

We propose FlashTTS, a low-latency LLM-based streaming TTS framework addressing the high latency bottleneck in real-time dialogue systems via a lagged multi-track architecture, parallel MTP, and an X-pred Mean Flow decoder. Its multi-track formulation natively processes streaming inputs, completely circumventing traditional sentence-level buffering. MTP and X-pred jointly accelerate acoustic generation, where the 2-NFE MTP-3 configuration optimally balances decoding speed and synthesis stability. Experiments demonstrate FlashTTS significantly improves streaming efficiency over robust baselines, reducing FTL to 60ms and FPL to 325ms while preserving competitive WER, strong zero-shot SIM, and positive subjective CMOS. Overall, FlashTTS successfully bridges the gap between high-fidelity speech synthesis and strict latency constraints in modern dialogue systems, providing a scalable foundation for real-time applications.

% \section{Discussion}
% While FlashTTS demonstrates significant efficacy in low-latency streaming synthesis, several avenues for future optimization remain. First, the current architecture relies on a relatively compact 0.5B parameter language model as its backbone. Scaling this foundation to larger model sizes presents a promising direction, as increased capacity typically enhances speculative capabilities, thereby potentially elevating the acceleration ratio achieved by the multi-token prediction (MTP) mechanism. Furthermore, the overall generation speed is intrinsically linked to the precision of the MTP module. Future work will focus on refining this module to improve prediction accuracy, which directly translates to reduced rejection rates and superior inference efficiency.

% In terms of acoustic modeling, we identify opportunities to further enhance the fidelity of the X-pred Mean Flow component. Improving the reconstruction quality of this module is crucial for capturing subtle acoustic details. Additionally, the current system operates by predicting Mel-spectrograms as intermediate representations. A significant frontier for future research involves exploring end-to-end paradigms that directly predict raw audio samples. We plan to investigate this approach by integrating aligned perceptual losses, aiming to eliminate the information loss associated with feature conversion and further elevate the naturalness of the synthesized speech.
\section{Generative AI Use Disclosure}
% % 必须披露生成式人工智能的使用程度。本部分内容可置于常规论文的第5或第6页，或长篇论文的第9或第10页。ISCA政策规定：所有（共同）作者必须对论文的工作和内容负责，并同意提交。任何生成式人工智能工具都不能作为论文的共同作者。它们可以用于编辑和润色稿件，但不应用于生成稿件的重要部分。
Generative AI tools are used in this work only for language editing, polishing, and formatting of the manuscript. They are not
used to generate any core content, research ideas, experimental
designs, results, or major textual parts of the paper. All scientific contributions, including model design, experiments, analysis, and conclusions, are completed by the authors.

\bibliographystyle{IEEEtran}
\bibliography{mybib}

\end{document}